\documentclass{article}
\usepackage{graphicx}
\usepackage{amsmath}
\usepackage{dcolumn}
\usepackage{bm}
\DeclareMathAlphabet{\mathpzc}{OT1}{pzc}{m}{it}

\newcommand{\matd}[1]{\frac{\text{D}{#1}}{\text{D}t}}

\usepackage{xcolor}
\usepackage[utf8]{inputenc}
\usepackage[T1]{fontenc}
\begin{document}
\title{On the shock change equations}
\author{
M. I. Radulescu\\
Department of Mechanical Engineering\\
University of Ottawa, Ottawa (ON) K1N 6N5 Canada\\
}

\date{\today}

\maketitle
\begin{abstract}
We revisit and derive the shock-change equations relating the dynamics of a shock wave with the partial derivatives describing the motion of a reactive fluid with general equation of state in a stream-tube with arbitrary area variation.  We specialize these to a perfect gas, in which we obtain all shock-change equations in closed form.  These are further simplified for strong shocks.  We discuss the general usefulness of these equations in problems of reactive compressible flow and in the development of intrinsic evolution equations for the shock, such as the approximations made by Whitham and Sharma.
\end{abstract}

\section{Introduction}
The shock-change equations provide the relations between variables characterizing the motion of a shock discontinuity (its location, shape and speed, and their derivatives) and the local partial derivatives of the flow variables on either side of the shock \cite{fickett}.  The shock-change equations can be very useful in treating problems of compressible flows, since they permit to link the smooth regions of the flow to the geometric descriptors of the shape and the kinematics of the shock surface (e.g., total curvature, velocity and acceleration) and the shock-jump conditions.  For example, in experimental work, the motion of the shock wave can be obtained accurately by high speed photography whereas other flow variables are difficult to measure.  Likewise, measurements can be made at fixed locations only.  The shock-change equations permit to reconstruct the flow field in the vicinity of the shocks or locations probed \cite{rakotoarison2019}.  In numerical work, for example, the shock surface can be treated as a computational boundary, along which the boundary conditions take the form of evolution equations to be solved simultaneously to the interior flow \cite{henrick, kasimov}.  They are also very useful in analytical work, particularly for reactive flow, since the reactivity in the gas along particle paths can be linked to the shock state and the shock dynamics \cite{lundstrom, eckett, vidal1999, radulescu2010}. More generally, the evolution of the shock can be coupled with the reactive gasdynamics in the smooth regions of the reaction zone \cite{clavin2002dynamics, short2001nonlinear, kasimov2005asymptotic, vidal2009}.  Generic local evolution equations for the shock dynamics can thus be obtained from certain simple approximations on (combinations of) partial derivatives evaluated at the shock, depending on the magnitudes of their coefficients \cite{ whitham1958propagation, sharma1994}, or from specific boundary conditions that define exactly one of these (combinations of) derivatives. 

In the present study, we wish to review and formulate these shock change equations, which appear in different forms and under different approximations (e.g., inert planar, cylindrical or spherical flow of perfect gases) in the litterature, in a unified way.  We choose as starting point the motion of a reactive compressible medium in a stream-tube with area variations in both time and space.  Following the procedure of Fickett and Davis \cite{fickett}, we formulate the shock-change equations for any partial derivative of interest for a general fluid.  We then specialize this for a perfect gas.  The expressions obtained for strong shocks in a perfect gas are sufficiently simple and useful in practice; we list the most useful results.  We conclude by illustrating the direct use of the shock-change equations as approximate evolution equations for predicting the shock dynamics given approximations for the partial derivatives of the flow in the smooth regions.  
  
\section{The reactive Euler equations}
We start with the general reactive Euler equations written for a stream-tube with varying area $A(x,t)$.  These describe the general motion of a compressible reactive fluid.
\begin{align}
\frac{1}{\rho}\matd{\rho}&=-\frac{\partial u}{\partial x} - \dot{\sigma}_A \label{eq1}\\
\rho \matd{u}&=-\frac{\partial p}{\partial x} \label{eq2}\\
\matd{p}&=c^2 \matd{\rho} + \rho c^2  \dot{\sigma} \label{eq3}
\end{align}
where $x$ and $t$ are the coordinate along the stream tube and time,  $\text{D}/\text{D}t=\partial/\partial t + u \partial/\partial x$ is the rate of change along the trajectory of a fluid particle.  The rate of strain of a fluid element in the transverse direction is
\begin{align}
\dot{\sigma}_A =  \matd{\ln A} \label{eq4}
\end{align}
while the thermicity $ \dot{\sigma}$ is 
\begin{equation}
\dot{\sigma}= - \frac{\rho}{c_p} \left(\frac{\partial v}{\partial T}\right)_{p, Y_i} \sum_{i=1}^N \left( \frac{\partial h}{\partial Y_i}\right)_{p, \rho, Y_{j,j \neq i}}\frac{D Y_i}{D t} \label{eq5}
\end{equation}
where $h$ is the mixture averaged enthalpy among the $N$ components and $Y_i$ is the mass fraction of component $i$.   Simpler expressions can be written for explicit equations of state \cite{fickett, Williams1985, Kao2008}.  
It is useful to eliminate $\matd{\rho}$ from \eqref{eq1} using \eqref{eq3}, such that compressible gasdynamics be described by variation in pressure and speed:
\begin{align}
\matd{p}=\rho c^2 \left( -\frac{\partial u}{\partial x} - \dot{\sigma}_A +  \dot{\sigma}\right) \label{eq6}
\end{align}
\section{Projection along a shock wave trajectory}
The partial differential equations \eqref{eq2} and \eqref{eq6} can be projected along arbitrary paths $x_{observ}(t)$ such that partial derivatives appearing in \eqref{eq1}-\eqref{eq3} can be expressed in terms of derivatives taken along the path $x_{observ}(t)$.  The speed of the observer being $S_{observ}=\dot{x}_{observ}(t)$, convective derivatives taken along the path $x_{observ}(t)$ are
\begin{align}
\left(\frac{d}{dt}\right)_{observ}=\frac{\partial}{\partial t}+S_{observ}\frac{\partial}{\partial x} \label{eq7}
\end{align}
such that
\begin{align}
\matd{}=\left(\frac{d}{dt}\right)_{observ} + (u-S_{observ})\frac{\partial}{\partial x} \label{eq8}
\end{align}
Here we are interested in projecting the governing equations along the motion of a thin shock wave (subscript $w$), such that weak solutions to the governing equations \eqref{eq1}-\eqref{eq3} (i.e., the Rankine Hugoniot jump equations) dictate the post wave state given the wave speed  $S_w$ and the state of the medium ahead of the wave.  Using \eqref{eq8}, our governing equations \eqref{eq2} and \eqref{eq6} become respectively:
\begin{align}
 \left(\frac{d u}{dt}\right)_{w} +(u-S_{w})\frac{\partial u}{\partial x}=-\frac{1}{\rho}\frac{\partial p}{\partial x} \label{eq9}\\
 \left(\frac{d p}{dt}\right)_{w} +(u-S_{w})\frac{\partial p}{\partial x}=\rho c^2 \left( -\frac{\partial u}{\partial x} - \dot{\sigma}_A +  \dot{\sigma}\right) \label{eq10}
\end{align}
Since we are following the shock, all the variables in equations \eqref{eq9} and \eqref{eq10} refer to the post-shock state. As such, the ratio $(du/dt)_w/(dp/dt)_w$ becomes the variation of particle speed with pressure along the shock Hugoniot, 
\begin{equation}
\left(\frac{du}{dp}\right)_H = \frac{\left(\frac{d u}{dt}\right)_{w}}{\left(\frac{d p}{dt}\right)_{w}} 
\end{equation}
the curve marking the loci of possible post shock states.  This is a property of the material's equation of state.  Solving for the two derivatives $\frac{\partial p}{\partial x}$ and $\frac{\partial u}{\partial x}$ from these two equations, we get:
\begin{align}
\frac{\partial u}{\partial x} =\eta^{-1}\left( \dot{\sigma} -\dot{\sigma}_A - \frac{1}{\rho c^2} \left( \frac{dp}{dt}\right)_{w}  \left( 1+\rho_0(S_w-u_0) \left(\frac{du}{dp}\right)_H \right)
\right) \label{eq11}\\
\frac{\partial p}{\partial x} =\rho_0( S_w-u_0)\eta^{-1}\left( \dot{\sigma} -\dot{\sigma}_A - \frac{1}{\rho c^2} \left( \frac{dp}{dt}\right)_{w}  \left( 1+ \frac{\rho_0(S_w-u_0)}{1-\eta} \left(\frac{du}{dp}\right)_H\right)
\right) \label{eq12}
\end{align}
where we have used the mass conservation across the shock wave $\rho_0 (S_w-u_0) = \rho (S_w-u)$, subscript 0 denotes the state ahead of the wave and the sonic parameter $\eta$ is defined as:
\begin{equation}
\eta=1-\left(\frac{S_w-u}{c}\right)^2 \label{eq13}
\end{equation}
Note that $S_w-u$ is the post-shock flow speed in the frame of reference of the shock wave and 
\begin{equation}
M=\frac{S_w-u}{c}\label{eq13_2}
\end{equation}
is its Mach number in that frame of reference. 

Equations \eqref{eq11} and \eqref{eq12} are the so-called shock change equations.  They give the relation between the shock wave acceleration (through the term $\left(dp/dt\right)_{w}$) with respect to the pressure or velocity gradients behind the wave and the influence of geometry and energy release in the post-shock state.  These equations specialized to planar, cylindrical or spherical motion have been derived by Fickett and Davis \cite{fickett}, the generalizations here are the arbitrary geometry, which can also be a function of time through the general term $\dot{\sigma}_A$ and arbitrary thermicity in any multi-component medium.  

In practice, analogous shock change equations can be formulated relating the shock motion to the flow acceleration behind the shock, for example in hydrodynamic stability problems.  From the momentum equation \eqref{eq2}, we obtain immediately:
\begin{equation}
a=\matd{u}=-\frac{1}{\rho}\frac{\partial p}{\partial x} \label{eq14}
\end{equation}
and its corresponding dependence on shock acceleration follows from \eqref{eq12}.

In reactive problems, the volumetric expansion of the gas behind the leading shock dictates the ignition along particle paths \cite{lundstrom, vidal1999, vidal2009, eckett, radulescu2018criticalN}.   From the continuity equation \eqref{eq1} and \eqref{eq11}, we get immediately
\begin{equation}
\frac{1}{\rho}\matd{\rho}=\eta^{-1}\left( -\dot{\sigma} +\dot{\sigma}_A(1-\eta) + \frac{1}{\rho c^2} \left( \frac{dp}{dt}\right)_{w}  \left( 1+\rho_0(S_w-u_0) \left(\frac{du}{dp}\right)_H \right)
\right) \label{eq15}
\end{equation}

In experiments, pressure measurements are performed at fixed locations, hence it is useful to relate local pressure changes with the shock dynamics.  We can write:
\begin{equation}
\frac{\partial p}{\partial t}=\left( \frac{dp}{dt}\right)_{w} - S_w \frac{\partial p}{\partial x}
\end{equation}
where the closed dependence on shock acceleration comes from \eqref{eq12}.   It is now evident that other derivatives of interest behind the shock, for example along characteristics, e.g., 
\begin{equation}
\frac{\partial p}{\partial t} + (u \pm c) \frac{\partial p}{\partial x} \label{eq16}
\end{equation}
can be easily written with manipulations of \eqref{eq1}-\eqref{eq3} and the two principal results given by \eqref{eq11} and \eqref{eq12} . 
    
\section{Hugoniot parametrized by $M_w$ for arbitrary media}
The right-hand-side of the shock change equations developed in the previous section can be written in terms of a single variable measured behind the shock and its time derivative, since all the other ones can be found from the Rankine-Hugoniot equations and the upstream state.  Although this choice is arbitrary, the normal flow speed relative to the shock $S_w-u_0 $ ahead of the shock, or its Mach number, i.e., 
\begin{equation}\label{eqRH1}
M_w=\frac{S_w-u_0}{c_0}
\end{equation}
can be used for this purpose as independent variable.  We can write:
\begin{equation}
\left( \frac{dp}{dt}\right)_{w}=\left( \frac{dM_w}{dt}\right)_{w} \left( \frac{dp}{dM_w}\right)_{H}=\dot{M}_w \left( \frac{dp}{dM_w}\right)_{H} \label{eq17}
\end{equation}
where $\left( dp/dM\right)_{H}$ is also a property of the shock Hugoniot.  Since we can also write:
\begin{align}
\left(\frac{du}{dp}\right)_H=\left(\frac{du}{dM_w}\right)_H \left(\frac{dM_w}{dp}\right)_H \label{eq18}
\end{align}
the right hand sides of all shock change equations listed can be re-written in terms of the wave Mach number, its rate of change, shock Hugoniot properties and the upstream state.  

For a perfect gas (see below) and its simple generalization as the Nobel-Abel Stiffened Gas \cite{radulescu2020compressible}, or for condesed media characterized by sufficiently simple equations of state (e.g., see Rabie and Wackerle \cite{rabie1978}), these expressions can be written in closed form.   For media characterized by more complex equations of state requiring the use of numerical evaluation, these can be easily evaluated numerically when the shock states can be parametrized by the shock speed (or its Mach number).  For example, numerical tools, such the Gordon and McBride's \textit{Chemical Equilibrium with Applications} \cite{CEA} , or the most recent Python implementation in the Shock and Detonation Toolbox \cite{Browne2006} for Cantera \cite{Cantera}, can determine the shock Hugoniot in mixtures of ideal gases with variable specific heats.  Rakotoarison et al.\ have recently used this numerical approach to reconstruct the experimental flow fields given the information on the shock dynamics \cite{rakotoarison2019}.   

\section{State and flow derivatives at a shock in a non-reactive perfect gas}
If one assumes a non-reactive shock wave, the shock jump conditions for the variables of interest are well known \cite{Whitham1974}. With the Mach number of the shock wave propagating with respect to the non-shocked material given by \eqref{eqRH1}, the shock jump equations are 
\begin{align}
\frac{u-u_{0}}{c_0}&=\frac{2\left(M_w^2-1\right)}{\left(\gamma+1\right)M_w}\label{eqRH2}\\
\frac{\rho}{\rho_0}&=\frac{\left(\gamma+1\right)M_w^2}{\left(\gamma-1\right)M_w^2+2}\label{eqRH3}\\
\frac{p-p_0}{p_0}&=\frac{2\gamma\left(M_w^2-1\right)}{\left(\gamma+1\right)}\label{eqRH4}
\end{align}
and the sound speed is $c^2=\gamma p/ \rho$.  The necessary derivatives in \eqref{eq17} and \eqref{eq18} with the Mach number can be evaluated.  After some straightforward algebra, the resulting shock change equations of interest are sufficiently simple.  Without the thermicity terms and specializing the problem to stream tubes varying with $x$ only, and noting that the shock wave curvature is $\kappa = d \ln A/dx$ we obtain:
\begin{equation}
\frac{\partial u}{\partial x} = \frac{
-2\left( (3M_w^2+1)\dot{M}_w(\gamma+1) + c_0\kappa(M_w^2-1)(1+\gamma (2M_w^2-1))\right)
}{M_w(M_w^2-1)(\gamma+1)^2} \label{RH5}
\end{equation}

\begin{multline}
\frac{\partial p}{\partial x} = 
-\frac{
2 \rho_0 c_0 \dot{M}_w(\gamma +1)  \left(2(2 \gamma -1) M_w^4+(\gamma +5) M^2-(\gamma-1)\right)
  }{(M_w^2-1)(2+M_w^2(\gamma-1))(\gamma+1)^2}\\
-\frac{
\rho_0 c_0 \kappa (M_w^2-1)(2 +  M_w^2 (\gamma-1)) (1 + \gamma(2 M_w^2-1  ))  }{(M_w^2-1)(2+M_w^2(\gamma-1))(\gamma+1)^2}
  \label{RH6}\\
\end{multline}

\begin{equation}
\frac{1}{\rho}\matd{\rho} =  \frac{
2\left( (3M_w^2+1)\dot{M}_w(\gamma+1) + c_0\kappa(M_w^2-1)(2+M_w^2(\gamma-1))\right)
}{M_w(M_w^2-1)(\gamma+1)^2} \label{RH7}
\end{equation}

Although the shock change equations derived above are useful in evaluating the flow derivatives given the shock geometry, these become more transparent for strong shocks, taken in the limit $M_w^2 \gg 1$.  These desired relations can be either obtained directly from \eqref{RH5}-\eqref{RH7} in the strong shock limit, or by starting with the equations \eqref{eq11}, \eqref{eq12} and \eqref{eq15}, and writing down the Rankine Hugoniot equations for a strong shock and evaluating the necessary derivatives along the Hugoniot.  Either way, the resulting shock change equations can be written in non-dimensional form as:  
\begin{equation}
\frac{S_w}{\dot{S}_w}\frac{\partial u}{\partial x} =-\frac{6}{\gamma+1} -\frac{4 \gamma}{(\gamma+1)^2}\frac{S_w^2 \kappa}{\dot{S}_w}\label{RH8}
\end{equation}

\begin{equation}
\frac{1}{\dot{S}_w}\frac{\partial u}{\partial t} =\frac{8}{\gamma+1} +\frac{4\gamma }{(\gamma+1)^2}\frac{S_w^2 \kappa}{\dot{S}_w}\label{RH8b}
\end{equation}

\begin{equation}
\frac{\left(\gamma-1\right)}{\rho_0 \dot{S}_w} \frac{\partial p}{\partial x} =
 -\frac{4 \left( 2 \gamma^2+\gamma-1 \right) }{(\gamma+1)^2}
 -\frac{4 \gamma \left(\gamma-1\right)}{(\gamma+1)^2} \frac{S_w^2 \kappa}{\dot{S}_w} \label{RH9}
\end{equation}

\begin{equation}
\frac{\left(\gamma-1\right)}{\rho_0 S_w \dot{S}_w}\frac{\partial p}{\partial t} = \frac{4 (3\gamma^2+\gamma-2) }{(\gamma+1)^2} 
+ \frac{4 \gamma(\gamma-1)^2}{(\gamma+1)^2 } \frac{S_w^2 \kappa}{\dot{S}_w}
  \label{RH9b}
\end{equation}

\begin{equation}
\frac{\left(\gamma-1\right)S_w}{\rho_0 \dot{S}_w}\matd{\rho} =  6+\frac{2(\gamma-1)}{(\gamma+1)}  \frac{S_w^2 \kappa}{\dot{S}_w}\label{RH10}
\end{equation}

\noindent where we have also taken $u_0=0$ for simplicity.  In the above set of shock-change equations, each one of the three-terms are non-dimensional, using the shock speed $S_w$ as characteristic velocity, its change time $(S_w/\dot{S}_w)$ as characteristic time and $\rho_0 S_w^2$ as characteristic pressure.  The first terms on the right hand sides measure the effect of non-steadiness on the partial derivatives, while the second term measures the effect of geometrical flow divergence.  These are written in such a way as to permit to evaluate their magnitude by inspection in the limit $\gamma \rightarrow 1$ discussed below, for example.

\section{Shock change equations as evolution equations for the shock}
Each of these shock change equations, or linear combination of these, can also be interpreted as an evolution equation for the shock, provided one, or a group of the terms appearing on the left hand side of these equations is known, can be modeled, or matched to an analysis of the flow behind the shock.   This step is of course problem dependent, and no unique prescription of the partial derivatives apply to all problems.  

The first step is to recognize that the sole term appearing on the right-hand-side of the these equations is simply
\begin{equation}
\frac{S_w^2 \kappa}{\dot{S}_w}=\frac{d (\ln A)}{d (\ln S_w)}
\end{equation}
Hence, by requiring that the left hand sides are approximately zero, each of the equations listed, or combinations of them, become evolution equations that yield a power law of the form $S_w \sim A^m$, where the power $m$ depends on the problem solved.  Among these, two useful approximations have been proposed for inert shocks.  

One physically based approximation is to assume that the changes of the pressure along the shock is approximately that along a particle path, a situation that applies in the Newtonian limit $\gamma\rightarrow 1$ of strong shocks, or the so-called \textit{snow-plow} approximation \cite{chernyi}. Since the shock density, $\rho/\rho_0 = (\gamma+1)/(\gamma-1)$, becomes infinite in this limit, the trajectory of a particle path coincides with the trajectory of the shock.  One can argue that both trajectories then measure approximately the same pressure changes.  This yields the requirement that    

\begin{equation}
\left( \frac{dp}{dt}\right)_{w} - \matd{p}=(S_w-u)\frac{\partial p}{\partial x} \simeq 0
  \label{RH11}
\end{equation}
This corresponds to a negligible pressure gradient behind the shock.  By inspection of  \eqref{RH9}, this yields 
\begin{equation}
S_w \sim A^{-\frac{\gamma(\gamma-1)}{(2\gamma-1)(\gamma+1)}}
  \label{RH12}
\end{equation}

The approximation of vanishing pressure gradient was assumed in the work of Sharma and co-workers, which verified favourably when compared to experiment and Whitham's approximation \cite{ridoux, ridouxphd}.  This truncation, however, when viewed in conjunction with \eqref{RH9} poses a difficulty in the limit $\gamma \rightarrow 1$, as the curvature would require to be much larger than the non-steady effects, making it a quasi-steady approximation.  

Another approximation is that of Whitham \cite{whitham1958propagation}, building on earlier work by Moeckel and Chisnell, which assumes that 
\begin{equation}
\left(\frac{1}{S_w}-\frac{1}{u+c}\right)  \left(\frac{\partial p}{\partial t} + \rho c\frac{ \partial u}{\partial t} \right)\simeq 0
  \label{RH13}
\end{equation}
at the shock.  This is what is known as Whitham's \textit{characteristic rule}, which states that a simple evolution equation for the shock can be obtained when derivatives along the shock approximate derivatives along the $C^+$ characteristics, i.e.,
\begin{equation}
\left( \frac{1}{u+c}\frac{\partial }{\partial t} + \frac{\partial }{\partial x} \right) \simeq \left( \frac{1}{S_w}\frac{\partial }{\partial t} + \frac{\partial }{\partial x}  \right)
  \label{RH14}
\end{equation}
for the $C^+$ characteristics:
\begin{align}
\left(\frac{1}{u+c}\frac{\partial  }{\partial t} +\frac{\partial }{\partial x}\right)p+\rho c \left(\frac{1}{u+c}\frac{\partial  }{\partial t} +\frac{\partial }{\partial x}\right) u = \frac{\rho c^2}{u+c} \frac{d (\ln A)}{dx}  \label{RH15}
\end{align}
obtainable directly from a linear combination of \eqref{eq2} and \eqref{eq6}. Using the approximation \eqref{RH14}, \eqref{RH15} is re-written as:  
\begin{align}
\left(d p \right)_w+\rho c \left( d u \right)_w = \frac{\rho c^2}{u+c} d( \ln A)  \label{RH16}
\end{align}
or 
\begin{align}
\left(\frac{d p}{d M_w}  \right)_H +\rho c \left(\frac{d u}{dM_w}  \right)_H = \frac{\rho c^2}{u+c} \frac{d( \ln A)}{dM_w}   \label{RH17}
\end{align}
This is Whitham's celebrated evolution equation for $A(M_w)$ once the terms appearing on the left hand side are evaluated from the shock Hugoniot given the relevant equation of state.  Clearly, the same evolution equation for $A(M_w)$ could have been obtained by substituting the shock change equations for each of the partial derivatives necessary directly into \eqref{RH13}.   After some algebra, in the limit of strong shocks, Whitham's truncation\cite {Whitham1974} yields:
\begin{equation}
S_w \sim A^{-\frac{1}{1+\frac{2}{\gamma}+\sqrt{\frac{2\gamma}{\gamma-1}}}}
\end{equation}

Note that neither Sharma's nor Whitham's truncations are exact.  Whitham's solution is known to work very well in implosion-type problems, particularly for the Guderley similarity solution (which is of the singular type, or of the second kind \cite{lee_2016}), better than Sharma's approximation \cite{Whitham1974, sharma1994, ridoux, ridouxphd}. This agreement illustrates that Whitham's local approximation works well in problems in which the rear boundary conditions affect the dynamics on time scales much longer than those associated with the local area changes affecting the dynamics of the shock.   An argument to this effect, although qualitative, was given by Whitham, although a rational argument is currently still lacking, in spite of the problem being nearly 70 years old!

Nevertheless, the limitation of both truncations is best illustrated in self-similar problems of the first kind \cite{lee_2016}, such as Taylor-Sedov blast waves.  In these problems, the shock decay is uniquely controlled by the rear boundary conditions through the conservation of total energy engulfed by the blast, which remains constant and sets the condition of self-similarity of the problem \cite{lee_2016}.  Not surprisingly, neither Sharma's, nor Whitham's truncations approximate the blast decay correctly in this case.   For example, the shock wave in Taylor-Sedov blast waves grows as $R \sim t^\frac{2}{j+3}$, where $j$ is 0, 1 or 2 for planar, cylindrical or spherical blast waves. Evaluating the shock speed and its acceleration from this law, with $\kappa=j/R$, we obtain:
\begin{equation}\label{eq:sedov}
\frac{S_w^2 \kappa}{\dot{S}_w}=-\frac{2j}{j+1}
\end{equation}	 
This expression takes on values of 0, -1 and -4/3 respectively for planar, cylindrical and spherical blast waves.  Clearly, the Taylor-Sedov decay law is incompatible with Sharma's law (and Whitham's law for that matter), which, by setting the left hand side of  \eqref{RH9} to zero yields 
\begin{equation}
\frac{S_w^2 \kappa}{\dot{S}_w}=-\frac{2 \gamma^2+\gamma -1 }{\gamma (\gamma-1)}
\end{equation}	
In the limit of $\gamma$ tending to unity, this term becomes infinite, clearly different from the order unity Taylor-Sedov self-similar solution!   The same argument can be reformulated by writing the evolution of the Taylor-Sedov blast shock with its area,
\begin{equation}
S_w \sim A^{-\frac{j+1}{2j}}
\end{equation}
The exponent of the surface area takes on a finite value in the limit of $\gamma$ tending to unity, whereas Sharma's and Whitham's truncations predict this exponent to be zero, signifying a shock with constant speed. 

In closing this section, it should be mentioned that a quantitative justification for Sharma's and Whitham's approximations are still lacking for generic problems of shock evolution.   It should have become clear by now that the shock change equations do \textit{not} permit to obtain this justification, as the influence of the rear boundary conditions and the time scales involved require careful analysis in each case.  Instead, the shock change equations only provide the closure on the shock dynamics once the derivatives of the flow behind the shock are determined or approximated!  Alternatively, it permits to obtain the properties of the flow, given the shock decay law.  Returning to self-similar Taylor-Sedov blast waves, for example, the shock change equations, since they are exact, permit to determine, for example, the expansion rate of a fluid element behind the shock via \eqref{RH10} and the decay law given by \eqref{eq:sedov}, information useful to model the reactivity of a gas particle undergoing expansion \cite{lundstrom, eckett, vidal1999, radulescu2010}.  It also permits to infer, for example, the importance of curvature caused expansion versus the one caused by non-steady shock motion in \eqref{RH10}.  This ratio becomes, for Taylor-Sedov blast waves, 
\begin{equation}
-\frac{1}{3}\frac{\gamma-1}{\gamma+1}\frac{2j}{j+1}
\end{equation} 
which tends to zero as $\gamma \rightarrow 1$, illustrating that the curvature in this case does not control the gas expansion \cite{eckett}.

\section{Conclusions}
The present survey provided a unified simple approach to the derivation of the shock change equations, first introduced by Fickett and Davis for planar, cylindrical and spherical shocks for general equation of state.  We generalized this approach to arbitrary rate of lateral strain of the flow.  We then derived the general shock change equations that apply for a perfect gas, further specializing them for strong shocks.  While the resulting shock change equations can be used as evolution equations in problems where the remaining partial derivative can be modeled or matched to the smooth flow, their utility is mainly for determining exactly the gradients of the flow given the shock dynamics, or vice-versa, provide the dynamics of the shock given the flow evolution in the interior. 

\section*{Acknowledgements} 
Discussions with Pierre Vidal from the Institut Pprime of the Ecole Nationale Supérieure de Mécanique et d'Aérotechnique are greatly acknowledged.   Financial support was provided by the Natural Sciences and Engineering Research Council of Canada (NSERC) through the Discovery Grant ”Predictability of detonation wave dynamics in gases: experiment and model development”.  The material developed in this paper was originally developed as Lecture Notes for a Gasdynamics course taught by the author at the University of Ottawa.  
\bibliography{references}
\bibliographystyle{IEEEtran}
\end{document}